\documentclass[twocolumn,showpacs,prl,amsmath,amssymb]{revtex4}
\usepackage{colordvi,epsfig,bm,amssymb}
\newcommand{\cW}{{\cal W}}

\newcommand{\rd} {\mathrm d}
\newcommand{\re} {\mathrm e}
\newcommand{\ri} {\mathrm i}

\newcommand{\vp} {{\bm p}}

\newcommand{\vu} {{\bm u}}
\newcommand{\vv} {{\bm v}}
\newcommand{\vw} {{\bm w}}

\newcommand{\nn}{\nonumber} 
\newcommand{\be}{\begin{equation}} 
\newcommand{\ee}{\end{equation}}  
\newcommand{\bea}{\begin{eqnarray}}
\newcommand{\eea}{\end{eqnarray}}
\begin{document}
\title{Spectra of Random Stochastic Matrices and Relaxation in Complex Systems}
\author{Reimer K\"uhn}
\affiliation{Department of Mathematics, King's College London, Strand, London WC2R 2LS, UK}
\date{\today}
\begin{abstract}
\noindent
We compute spectra of large stochastic matrices $W$, defined on sparse random 
graphs, where edges $(i,j)$ of the graph are given positive random weights $W_{ij}
>0$ in such a fashion that column sums are normalized to one. We 
compute spectra of such matrices both in the thermodynamic limit, and for single 
large instances. The structure of the graphs and the distribution of the non-zero 
edge weights $W_{ij}$ are  largely arbitrary, as  long  as the mean vertex degree 
remains finite in the thermodynamic limit and the $W_{ij}$ satisfy a detailed balance 
condition. Knowing the spectra of stochastic matrices is tantamount to knowing the 
complete spectrum  of relaxation times  of stochastic  processes described by them, 
so  our results  should have  many interesting  applications for the description of 
relaxation in complex systems. Our approach allows to disentangle contributions to 
the spectral density related to extended and localized states, respectively, allowing
to differentiate between time-scales associated with transport processes and those 
associated with the dynamics of local rearrangements.
\end{abstract}


\maketitle


There are numerous processes, both natural and artificial, which can be understood in  
terms of random walks on complex networks \cite{Lov93, NohRie04, Bonav+14}, including 
the spread of diseases in social networks \cite{More+02, New02}, the transmission 
of information in communication networks (e.g. \cite{Spyr08}), search algorithms 
\cite{BriPag98, Adamic+01}, the out-of-equilibrium dynamics of glassy systems at low 
temperatures as described in terms of hopping between long-lived states in state space
\cite{Bou92, BarrMez95, Mor+11}, the dynamics of major conformational changes in 
macro-molecules \cite{NoeFi08}, or cell-signalling through protein-protein interaction 
networks \cite{Tesch+14}, to name but a few. For reviews that cover several of these 
topics, see e.g. \cite{New03, BarrBarthVes08, Dorog+08}.

The purpose of the present letter is to use random matrix theory to contribute to the 
understanding of systems of this type. We compute spectra of transition matrices 
for discrete Markov chains describing stochastic dynamics in complex systems. We  
construct these in terms of sparse random graphs in such a way that an edge $(i,j)$ in 
a graph corresponds to a possible transition $j\to i$, with the edge weight $W_{ij}>0$ 
quantifying the associated transitions probability, requiring $\sum_i W_{ij} = 1$ for 
all $j$. We are interested in the limit, where the number $N$ of possible states becomes 
large, with the average number of possible transitions at each state remaining finite 
in the thermodynamic limit ($N\to\infty$).

Given a time-dependent probability vector $\vp(t)=(p_i(t))$, we have an evolution 
equation of the form
\be
\vp(t+1) = W \vp(t)\ .
\label{dynp}
\ee
The condition $W_{ij}\ge 0$ for all $(i,j)$ and the column sum constraint together entail 
that the spectrum of $W$ is contained in the unit disc of the complex plane, $\sigma(W) 
\subseteq \{z; |z| \le 1\}$. If $W$ satisfies a detailed balance condition with an 
equilibrium distribution, $p_i = p_i^{\rm eq}$, such that $W_{ij} p_j = W_{ji} p_i$ for 
all pairs $(i,j)$, then $W$ can be symmetrized by a similarity transformation --- 
$\cW_{ij} = p_i^{-1/2} W_{ij} p_j^{1/2} = \cW_{ji}$ --- implying that the spectrum of 
$W$ is real, and $\sigma(W) \subseteq [-1,1]$.

Our main interest here is the relation between eigenvalues of $W$ and relaxation times of 
the Markov chain it describes. It is easily understood by following the evolution of an 
initial probability vector $\vp(0)$ over $t$ time steps, i.e. by considering $\vp(t) = W^t 
\vp(0)$. Using a spectral decomposition of $W$, and assuming the system to be irreducible 
and free of cycles, one obtains 
\be
\vp(t) = \vp^{\rm eq} + \sum_{\alpha (\ne 1)} \lambda_\alpha^t ~ \vv_\alpha ~
\big(\vw_\alpha , \vp(0)\big)
\label{p-evol}
\ee
where we have used that $1=\lambda_1 > |\lambda_\alpha|$ for $\alpha\ne 1$, given the 
assumptions \cite{Gantmacher59}, and where $\vv_\alpha$ and $\vw_\alpha$ denote the right 
and left eigenvectors of $W$, respectively, with $\vv_1= \vp^{\rm eq}$, and $\vw_1=(1,\dots,1)$. 
Eq. (\ref{p-evol}) allows to relate relaxation times of the system to eigenvalues of $W$ via 
$\tau_\alpha = -1/\ln|\lambda_\alpha|$ for $\alpha \ne 1$.

We construct random stochastic matrices in terms of unnormalized transition matrices
$\Gamma = (\Gamma_{ij}) = (c_{ij} K_{ij})$, with connectivity matrix elements $c_{ij} \in 
\{0,1\}$ (and $c_{ii}=0$) specifying the network structure of possible transitions, and 
positive edge weights $K_{ij}>0$, and setting
$
W_{ij} = \Gamma_{ij}/\Gamma_{j}
$
if $\Gamma_{j} \equiv \sum_i \Gamma_{ij} \ne 0$, and $W_{ii}= 1$ for isolated sites for which 
$\Gamma_{i} = 0$. The present investigation will be restricted to the case where $W$ satisfies
a detailed balance condition, and can thus be symmetrized by a similarity transformation, as
discussed above. The spectrum of fully connected matrices of this type was shown to converge
to a semi-circular law \cite{Bord+08} in the large system limit, and to a circular law, if the 
detailed balance condition is dropped \cite{Bord+12}. Asymptotic results related to the circular
law were obtained for Erd\"os-Renyi graphs with mean connectivity diverging in the thermodynamic 
limit \cite{Bord+14}. For some recent related results concerning spectra of graph Laplacians, we 
refer to \cite{Grab+12, Peix13, Zhang+14}.

We follow \cite{EdwJon76} and express the spectral density $\rho_W(\lambda)$ of the stochastic 
matrix $W$ in terms of a derivative 
\be
\rho_W(\lambda) = - \lim_{\varepsilon\to 0} \frac{2}{\pi N}\,{\rm Im}\, 
\frac{\partial}{\partial \lambda}\, \log Z_W(\lambda)\ ,
\label{rhoW}
\ee
of the logarithm of a Gaussian integral
\be
Z_W(\lambda) = \int \prod_{i=1}^N \frac{\rd u_i}{\sqrt{2\pi/\ri}}~
\exp\big\{-\ri H_W(\lambda_{\varepsilon},\vu)\big\}\ 
\ee
defined in terms of the quadratic form
\be
H_W(\lambda_{\varepsilon},\vu) =\frac{1}{2}\sum_{i,j}  \big(\lambda_{\varepsilon} 
\delta_{ij} - \cW_{ij}\big)\ u_i u_j\ ,
\label{HW}
\ee
with $\lambda_{\varepsilon}= \lambda - \ri {\varepsilon}$. Here, $\cW$ is the symmetrized 
version of $W$, obtained via a similarity transform that involves the equilibrium 
distribution $\vp^{\rm eq}$ as discussed above. The representation (\ref{rhoW}) allows 
to interpret the spectral density as a sum over single site variances
\be
\rho_W(\lambda) =  {\rm Re}\,\frac{1}{\pi N} \sum_i 
\langle u_i^2 \rangle 
\label{rhoW-ss}
\ee
of the complex Gaussian measure
\be
P_W(\vu) = \frac{1}{Z_W} \re^{-\ri H_W(\lambda_{\varepsilon},\vu)}\ . 
\label{PW}
\ee
Here and in what follows we shall omit explicitly writing the $\lim_{\varepsilon\to 0}$,
and take it to be understood.

In the thermodynamic limit, the spectral density is expected to be non-random and is 
obtained by averaging Eq. (\ref{rhoW}) over the matrix ensemble in question, using the 
replica method to perform averages as proposed in \cite{EdwJon76}, and taking the limit 
$N\to\infty$. Methods developed in \cite{Ku08} can be used to efficiently deal with the 
sparsity of the ensemble of matrices considered in the present letter. Alternatively, 
one can analyse single large instances using a cavity approach proposed in \cite{Rog+08} 
to obtain the single instance spectral density in terms of variances of single-site 
marginals. In the thermodynamic limit, recursion relations for the cavity-variances 
obtained within that approach can be interpreted as stochastic recursions, allowing to 
formulate self-consistency relations for their distributions, which turn out to be 
equivalent to those obtained using replica. This is the approach we shall briefly 
outline in what follows.

In order not to overburden the present exposition with technicalities, we shall 
restrict our attention here to cases where the unnormalized transition matrix 
$\Gamma$ is symmetric, in which case the symmetrized normalized Markov matrix is 
of the form
\be
\cW_{ij} = \frac{\Gamma_{ij}}{\sqrt{\Gamma_i \Gamma_j}}
\ee
for $\Gamma_{ij} >0$, hence $\Gamma_i>0$ and $\Gamma_j>0$, and $\cW_{ii}=1$ for 
isolated sites.

To obtain the single-site marginals of (\ref{PW}) required to evaluate $\rho_W(\lambda)$ 
according to (\ref{rhoW-ss}), we distinguish between single-site marginals on isolated sites, 
which are of the form $P^{\rm is}_i(u_i) \propto  \re^{-\frac{\ri}{2}(\lambda_{\varepsilon}-1)\,
u_i^2}$, and those for sites that are not isolated. On the latter, we perform a transformation 
of variables, $\frac{u_i}{\sqrt{\Gamma_i}} \to u_i$. In terms of the transformed variables, 
we have
\be
\rho_W(\lambda) = p_N(0) \delta(\lambda -1) +
{\rm Re}\,\frac{1}{\pi N} \sum_{i} \Gamma_i \langle u_i^2 \rangle\ , 
\label{rhoW-tr}
\ee
with $p_N(0)=\frac{N^{\rm is}}{N}$ denoting the fraction of isolated sites, and only 
non-isolated sites with $\Gamma_i>0$ contributing to the second sum.

On a locally tree-like graph a marginal $P_i(u_i)$ of a (transformed) variable on a 
non-isolated site can be expressed in terms of cavity marginals $P_j^{(i)}(u_j)$ on sites 
in the neighbourhood $\partial i$ of $i$ as
$$
P_i(u_i) \propto \re^{-\frac{\ri}{2}\Gamma_i \lambda_{\varepsilon}\,u_i^2} \prod_{j\in 
\partial i}\, \int \rd u_j \, \re^{\ri K_{ij} u_i u_j } P_j^{(i)}(u_j)\ .
$$
The cavity marginals satisfy a set of self consistency equations 
$$
P_j^{(i)}(u_j) \propto \re^{-\frac{\ri}{2}\Gamma_j\lambda_{\varepsilon}\,  u_j^2}
\prod_{\ell\in \partial j\setminus i}\,\int \rd u_\ell \, \re^{\ri K_{j\ell} u_j u_\ell } 
P_\ell^{(j)}(u_\ell)\ .
$$
These relations are exact on trees; for finitely connected random graphs they become asymptotically 
exact in the thermodynamic limit. They are solved \cite{Rog+08} by complex Gaussians of the form
$$
P_j^{(i)}(u_j) = \sqrt{\omega_j^{(i)}/2\pi} \, \exp\Big\{-\frac{1}{2}\omega_j^{(i)} u_j^2\Big\}\ ,
$$
with ${\rm Re}\,\omega_j^{(i)} \ge 0$, entailing that the inverse cavity variances satisfy the 
self-consistency equations
\be
\omega_j^{(i)} = \ri \lambda_{\varepsilon}\, \Gamma_j +
\sum_{\ell\in \partial j\setminus i} \frac{K_{j\ell}^2}{\omega_\ell^{(j)}}\ .
\label{cav-var}
\ee
These can be solved iteratively on large single instance. Single-site marginals, too, will 
be Gaussian with inverse variances expressed in terms of solutions of (\ref{cav-var}) as
$\omega_i = \ri \lambda_{\varepsilon}\, \Gamma_i +\sum_{j\in \partial i}K_{ij}^2/\omega_j^{(i)}$.
In terms of these inverse variances of single-site marginals then, we have
\be
\rho_W(\lambda) = p_N(0) \delta(\lambda -1) + {\rm Re}\,\frac{1}{\pi N} \sum_{i} 
\frac{\Gamma_i}{\omega_i}\ .
\label{rhoW-G}
\ee
Specializing to the case of unbiased random walk, we have $\Gamma_{ij}=c_{ij}$, hence 
$\Gamma_j = k_j$ and $\cW_{ij}=\frac{c_{ij}}{\sqrt{k_i k_j}}$ for non-isolated sites, 
where $k_i$ and $k_j$ are degrees of vertices $i$ and $j$. In this case, Eqs. 
(\ref{cav-var})-(\ref{rhoW-G}) readily lend themselves for averaging over a graph-ensemble
in the thermodynamic limit, giving rise to a recursion for a probability density function
$\pi(\omega)$ for inverse cavity variances of the form
\be
\pi(\omega) = \sum_{k\ge 1} p(k) \frac{k}{c} \int \prod_{\nu=1}^{k-1} \rd 
   \pi(\omega_\nu)\ \delta(\omega - \Omega_{k-1})
\label{eq:pi}
\ee
in which $p(k)$ is the degree distribution (thus $p(k) k/c$ the probability of an edge to be 
connected to a site with degree $k$, and
$
\Omega_{k-1}=\Omega(\{\omega_\nu\}_{\nu=1}^{k-1}) = \ri \lambda_\varepsilon k + 
\sum_{\nu=1}^{k-1} \frac{1}{\omega_\nu}\ .
$
In terms of the solution of (\ref{eq:pi}), one obtains the spectral density of $W$ for a 
random graph with degree distribution $p(k)$ as
\bea
\rho(\lambda)&=& p(0) \delta(\lambda -1)\nn\\
& & \hspace{-4mm} + \frac{1}{\pi}\,\mbox{Re} \,\sum_{k\ge 1} p(k) \int \prod_{\nu=1}^{k} 
\rd \pi(\omega_\nu)\ \frac{k}{\Omega_k}\  .
\label{eq:Dos}
\eea
Contributions related to extended and localized states can be identified as explained in
\cite{Ku08}. The same results have been obtained within a replica approach \cite{Ku14b}. 

\begin{figure}
\epsfig{file=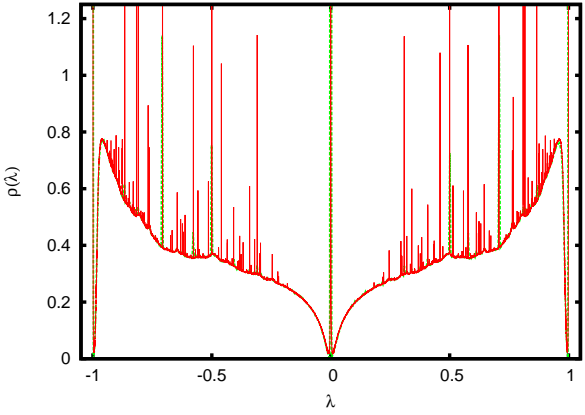, width=0.48\textwidth}
\caption{(Colour online) Spectral density of the transition matrix for an unbiased random 
walk on an Erd\"os-Renyi graph of mean coordination $c=2$, comparing results of numerical 
diagonalization of an ensemble of $1000 \times 1000$ matrices (green dashed curve) and 
analytic results obtained via population dynamics (red full curve).}
\end{figure}

For Markov processes other than the unbiased random walk, straightforward averaging of
cavity recursions over the ensemble of Markov matrices is prevented by the fact that the 
$K_{j\ell}$ in (\ref{cav-var}) are not independent due to column sum constraints, in
a way that extends beyond degree. In order to deal with this issue we return to the Gaussian
integral in terms of which the problem was originally formulated, and rewrite the quadratic
form (using transformed variables on non-isolated sites) as $H_W\!\!=\!\!\frac{1}{2}
\sum^{\rm i s}_{i}  (\lambda_{\varepsilon} -1) u_i^2 + \frac{1}{2}\sum_{i,j} c_{ij} 
\big[\frac{1}{2}\lambda_{\varepsilon} K_{ij} (u_i^2 +u_j^2) - K_{ij}\ u_i u_j\big]$.

Using this setup, one easily obtains the following reformulated recursion for inverse variances
of cavity marginals
\be
\omega_j^{(i)} = \sum_{\ell\in \partial j\setminus i}\Bigg(
\ri \lambda_{\varepsilon} K_{j\ell} + \frac{K_{j\ell}^2}{\omega_\ell^{(j)}+ 
\ri \lambda_{\varepsilon} K_{j\ell}}\Bigg)\ .
\label{cav-rec}
\ee
This version allows ensemble averaging, giving rise to the self-consistency equation
\be
\pi(\omega) = \sum_{k\ge 1} p(k) \frac{k}{c} \int \prod_{\nu=1}^{k-1} \rd \pi(\omega_\nu)\,
\Big\langle \delta(\omega- \Omega_{k-1})\Big\rangle_{\{K_\nu\}}
\label{piom}
\ee
with now
\be
\Omega_{k-1}=\sum_{\nu=1}^{k-1} \Bigg(
\ri \lambda_{\varepsilon} K_{\nu} + \frac{K_{\nu}^2}{\omega_\nu+ 
\ri \lambda_{\varepsilon} K_{\nu}}\Bigg)\ ,
\label{Om}
\ee
which is efficiently solved using a population dynamics algorithm. In terms of its solution, 
the spectral density in the thermodynamic limit is given by
\bea
\rho(\lambda) &=& p(0) \delta(\lambda - 1)\nn\\
& &\hspace{-4mm} + \frac{1}{\pi}\,\mbox{Re}\, \,\sum_{k\ge 1} 
p(k) \int \prod_{\nu=1}^{k} \rd \pi(\omega_\nu)\ 
\Bigg\langle\frac{\sum_{\nu=1}^k K_\nu}{\Omega_k}\Bigg\rangle_{\{K_\nu\}}\ .
\label{eq:StDos}
\eea
Fig. 1 shows the spectrum of the transition matrix for an unbiased random walk on an
Erd\"os-Renyi graph of mean connectivity $c=\langle k\rangle=2$, comparing results 
obtained from (\ref{eq:pi})-(\ref{eq:Dos}) for the thermodynamic limit with simulations 
averaged over 5000 realizations of $1000\times1000$ matrices, showing excellent agreement 
except that our population dynamics algorithm picks up many more of the localized states 
which appear as $\delta$-peaks in the diagram. A zoom into the $\lambda \lesssim 1$ region 
(not shown) would reveal a mobility edge at $\lambda_c \simeq 0.986$ and a Lifshitz-type
tail of eigenvalues at $\lambda\ge\lambda_c$ corresponding to a band localized states.

\begin{figure}
\epsfig{file=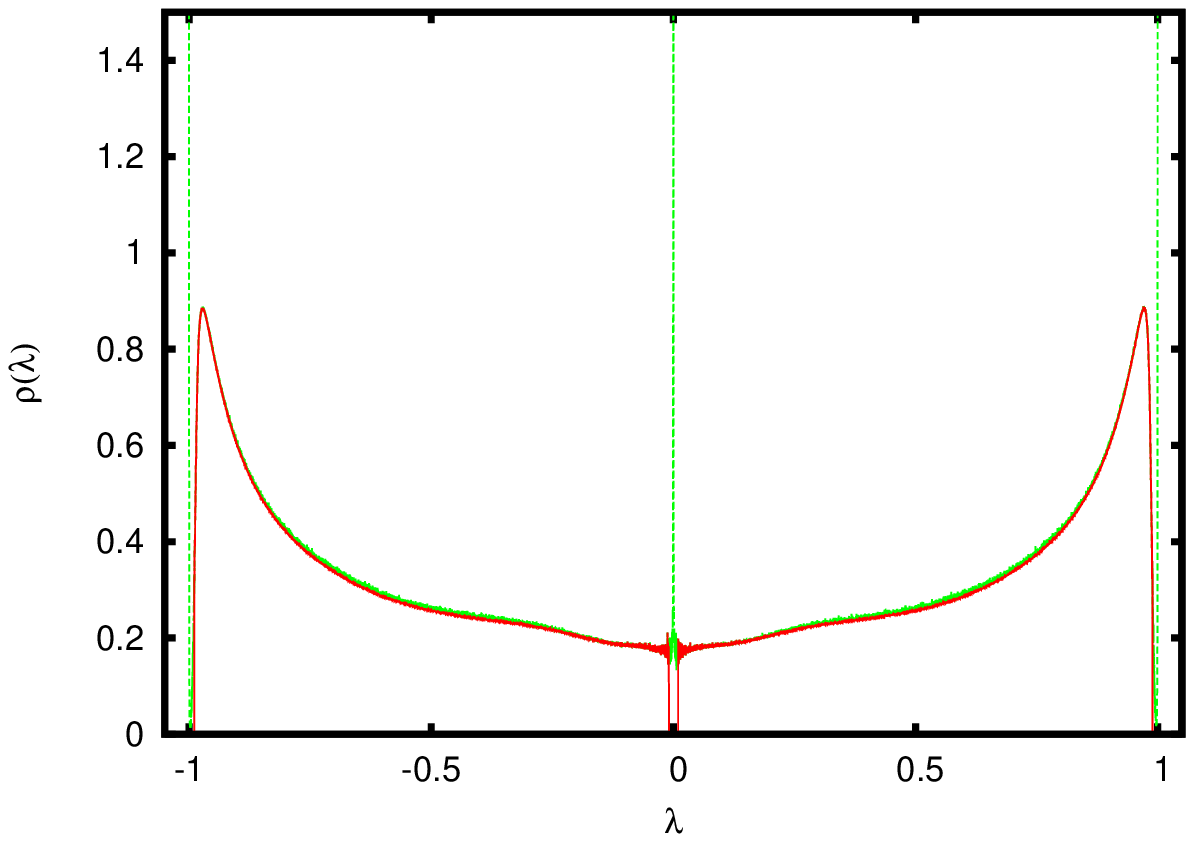, width=0.24\textwidth}\hfil
\epsfig{file=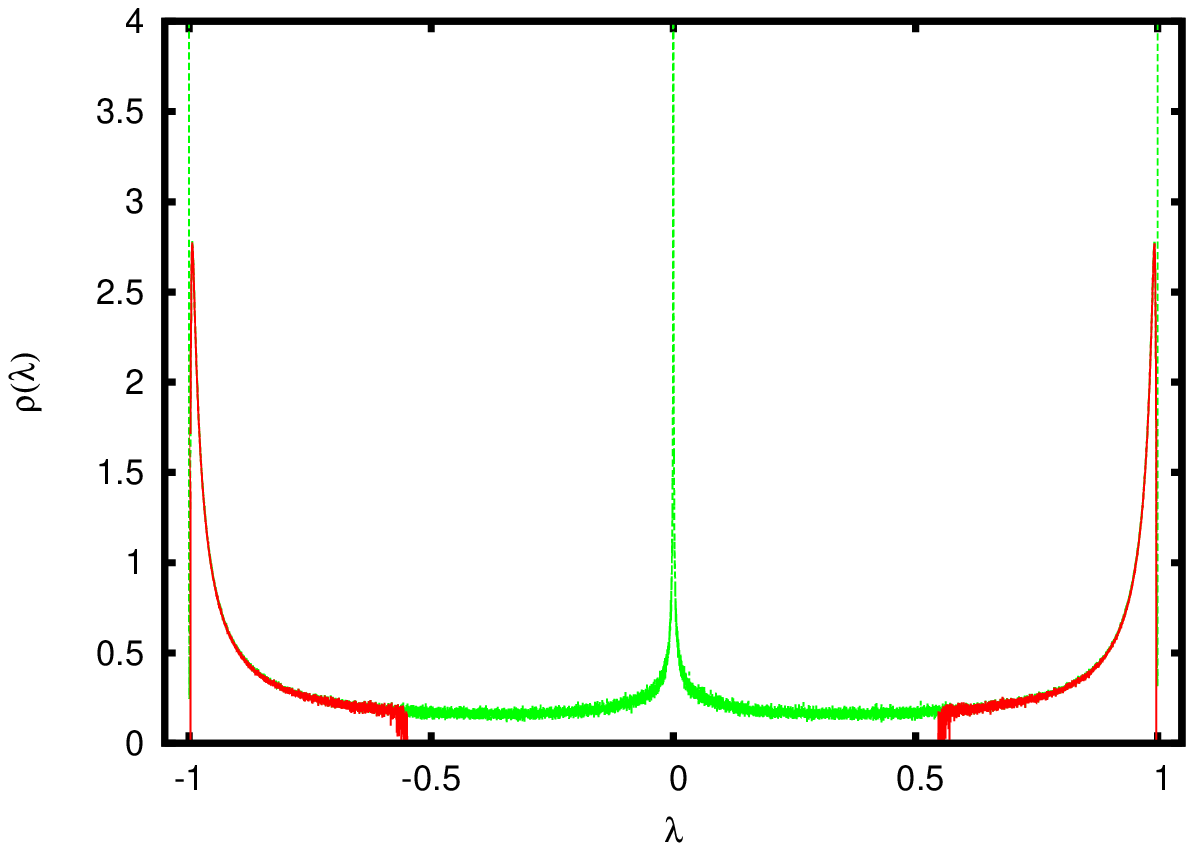, width=0.24\textwidth}
\caption{(Colour online) Spectral density of the transition matrix  with Kramers transition rates 
on an Erd\"os-Renyi graph of mean coordination $c=2$; shown are analytic results obtained via population 
dynamics, separately for the total density of states (green dashed curve) and the density of extended 
states (red). First panel: $\beta=2$; Second panel: $\beta=5$. (Note the different vertical scales.)}
\end{figure}

In Fig. 2 we present results for systems with unnormalized transition matrix elements taking the form of 
Kramers transition rates $\Gamma_{ij}=c_{ij}\re^{-\beta (V_{ij}-E_j)}$, with barrier heights $V_{ij}$ 
randomly and uniformly chosen in [0,1]; the distribution of initial energies in this case is arbitrary, as 
initial energies cancel in properly normalized stochastic matrices, so that $W_{ij} = c_{ij}\re^{-\beta V_{ij}}/
\sum_i c_{ij}\re^{-\beta V_{ij}}$. Systems of this type were studied within a heterogeneous mean-field 
approximation to dynamics in \cite{Mor+11}, generalizing earlier work \cite{Bou92, BarrMez95} to include 
barrier height distributions and incompletely connected networks of traps. Two aspects are particularly 
notable: (i) as $\beta$ is increased the spectral density gives more weight to regions near $\lambda=\pm 1$, 
i.e. to slow modes; (ii) the narrow region of localized states near $\lambda = 0$ broadens considerably, 
as $\beta$ is increased from 2 to 5, implying that many more modes have become localized. 
Once more, we found excellent agreement with simulation results \cite{Ku14b}.

For the unbiased random walk problem on a regular random graph with $p(k)=\delta_{k,c}$, Eqs (\ref{eq:pi}) 
are solved by a $\delta$-function, $\pi(\omega)= \delta(\omega-\bar \omega)$, giving rise to a quadratic 
self-consistency equation for $\bar\omega$; its solution, when inserted into (\ref{eq:Dos}), allows to 
obtain a closed-form expression for the spectral density
\be
\rho(\lambda) = \frac{c}{2\pi}\, \frac{\sqrt{4 \frac{c-1}{c^2} - \lambda^2}}{1-\lambda^2}\ ,
\label{KeMcK}
\ee 
which is readily recognised as a variant of the Kesten-McKay distribution \cite{McKay81}, adapted to 
capture the spectral problem of the Markov transition matrix for an unbiased random walk on random regular
graphs. The same result is found to provide an accurate approximate description for Erd\"os-Renyi random  
graphs at large mean degree $c$, which becomes asymptotically exact as $c\to\infty$, where (\ref{KeMcK}) 
approaches a semicircular law. An analogous line of reasoning allows to obtain the spectral density for 
more general Markov matrices on Erd\"os-Renyi and random regular random graphs in the large $c$ limit,
viz. the semi-circular law
\be
\rho(\lambda) =\frac{c}{2\pi} \frac{\langle K\rangle^2}{ \langle K^2\rangle} 
\sqrt{\frac{4\langle K^2\rangle}{c\langle K\rangle^2}-\lambda^2}\ .
\ee
This expression is invariant under rescaling of the edge weights $K_{ij}$, as it should, because $K$ 
scales are immaterial in normalized Markov transition matrices.

In summary, we computed spectra of random stochastic matrices defined in terms of random graphs, assuming 
that they satisfy a detailed balance condition. Of particular relevance is the possible appearance of 
localized states in such systems. Referring to Eq. (\ref{p-evol}), one can indeed argue that most modes 
corresponding to localized states will not contribute to the relaxation dynamics, if initial conditions 
are themselves localized, an issue we have not seen systematically investigated in the literature. Further 
details on several of the issues which could be just touched upon in the present letter will be provided 
in a forthcoming paper \cite{Ku14b}. We expect our methods and results to be of interest for the study 
of a broad range of relaxation phenomena in complex systems.\\[-20pt]
\bibliography{../../MyBib}
\end{document}